\documentclass[aps,prl,preprint,groupedaddress,floatfix]{revtex4}
\usepackage{graphicx}
\usepackage{epsfig}
\usepackage{amsmath}

\begin{document}
\title{Exchange bias of mu-metal thin films}

\author{H. F. Kirby}
\affiliation{Department of Physics, Center for Integrated Functional Materials,
University of South Florida, 4202 East Fowler Avenue, Tampa, FL 33620, USA}
\author{T. M. Eggers}
\affiliation{Department of Physics, Center for Integrated Functional Materials,
University of South Florida, 4202 East Fowler Avenue, Tampa, FL 33620, USA}
\author{P. B. Jayathilaka}
\affiliation{Department of Physics, Center for Integrated Functional Materials,
University of South Florida, 4202 East Fowler Avenue, Tampa, FL 33620, USA}
\author{S. M. Campbell}
\affiliation{Department of Physics, Center for Integrated Functional Materials,
University of South Florida, 4202 East Fowler Avenue, Tampa, FL 33620, USA}
\author{Casey~W.~Miller}
\affiliation{Department of Physics, Center for Integrated Functional Materials,
University of South Florida, 4202 East Fowler Avenue, Tampa, FL 33620, USA}
\email{millercw@usf.edu}

\begin{abstract}

The exchange bias of the soft ferromagnet mu-metal, Ni$_{77}$Fe$_{14}$Cu$_{5}$Mo$_{4}$, with the metallic antiferromagnet Fe$_{50}$Mn$_{50}$ has been studied as a function of ferromagnet thickness and buffer layer material. Mu-metal exhibits classic exchange bias behavior: the exchange bias ($H_{EB}$) and coercive fields scale inversely with the ferromagnet's thickness, with $H_{EB}$ varying as the cosine of the in-plane applied field angle.  While the exchange bias, coercivity, and exchange energy are greatest when the buffer layer material is (111) oriented Cu, amorphous Ta buffers allow the mu-metal to retain more of its soft magnetic character.  The ability to preserve soft ferromagnetic behavior in an exchange biased heterostructure may be useful for low field sensing and other device applications.

\end{abstract}
 \maketitle
\newpage
\section{Keywords}
exchange bias; mu-metal; Ni-Fe-Cu-Mo; soft ferromagnets

\section{Introduction}
Exchange bias is a phenomenon related to the interfacial exchange interaction between two ordered magnetic materials \cite{EBreviewIvan, josep-nanoeb-review}.  Observed primarily in structures composed of ferromagnet/antiferromagnet (FM/AF) interfaces (e.g., thin film heterostructures and nanoparticles), exchange bias manifests itself as a unidirectional magnetic anisotropy that shifts the hysteresis loop along the field axis by some amount known as the exchange bias, $H_{EB}$.  The ferromagnet has a unique magnetization at zero field when  $H_{EB}$ exceeds the saturation field, which allows simple FM/AF bilayers to serve as a magnetic reference for spintronics devices \cite{SpintronicsFundamentals}.   In fact, there are many potential applications of exchange bias because $H_{EB}$ is a function of many experimentally controllable parameters, including, but not limited to: ferromagnet and antiferromagnet thickness; temperature; interfacial structure and roughness; and grain size \cite{Bolon200754}.

In this work, we focus on inducing exchange bias in Ni$_{77}$Fe$_{14}$Cu$_{5}$Mo$_{4}$, which is sometimes referred to as  mu-metal or conetic. A member of the Permalloy family, this material has large permeability and saturation magnetization, and offers nearly zero magnetostriction and nearly zero magnetocrystalline anisotropy \cite{Egelhoff200690,jr.:013921}.  Introducing a unidirectional anisotropy via exchange bias in soft magnetic materials could be a useful for introducing additional control over phenomena and sensors such as giant magneto-impedance (GMI)  \cite{garcia:232501}.  Bulk mu-metal has been shown to have a large GMI ratio (300\%) and a correspondingly high sensitivity (20\%/Oe)  \cite{Nie1999285,Cho200051}, but its exchange bias properties have not been reported.  Another potential area for impact is exchange spring system that combine materials with perpendicular magnetic anisotropy with soft ferromagnet layers with in plane anisotropy.  This leads to structures whose magnetization has an out of plane tilt angle that is tunable by the thickness of the soft ferromagnet \cite{nguyen:172502}.  Such structures are being explored for spin transfer torque devices \cite{Ralph20081190}.

\section{Materials and Methods}
\begin{figure}
\begin{center}
\includegraphics[width=3.3in]{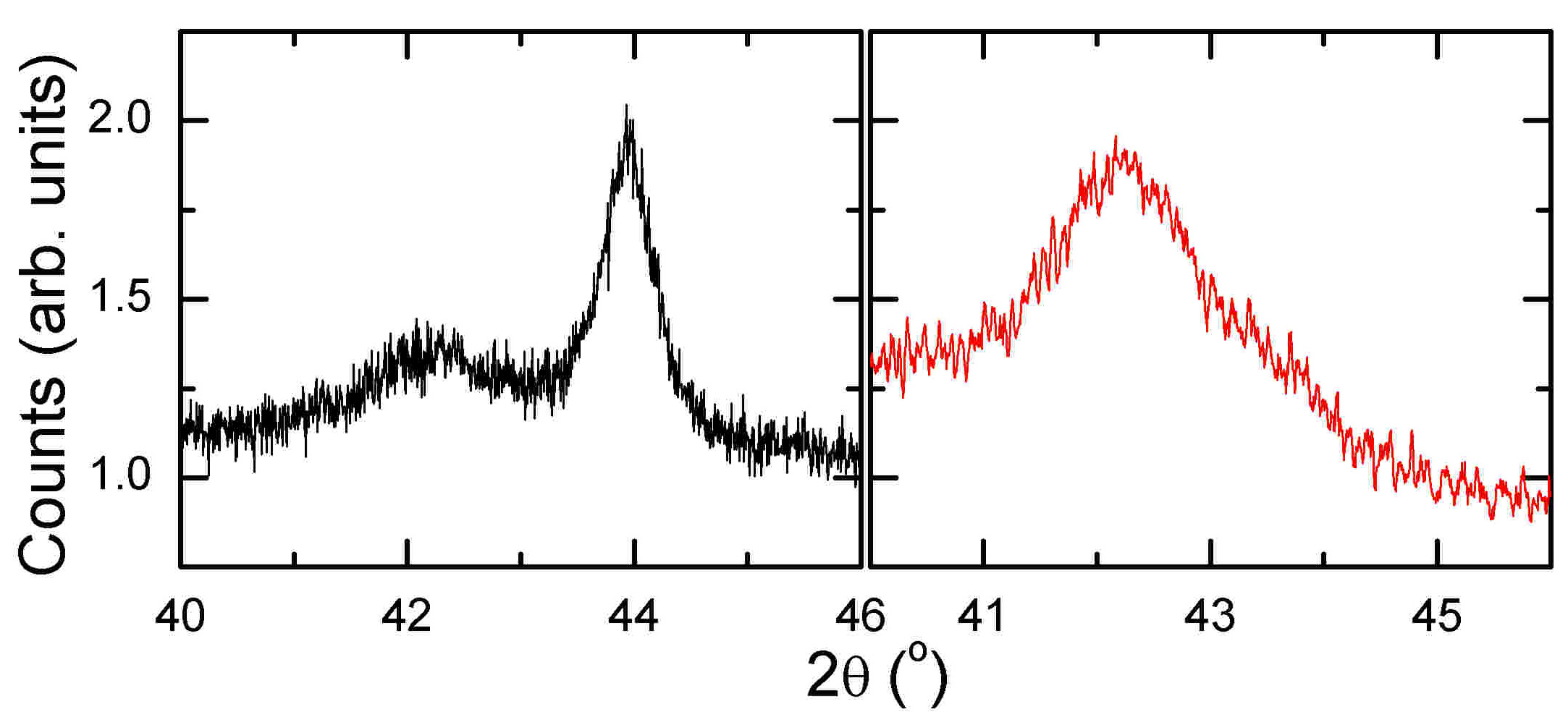}
     \caption{\small{(Color online) X-ray diffraction results show that the 50~$\textrm{\AA}$ Ta buffer (left) leads to more coherent (111) mu-metal texture than the 300~$\textrm{\AA}$ Cu buffer (right).\\
     }} \label{xrd}
\end{center}
\end{figure}We investigated how the magnetic properties of three sets of Ni$_{77}$Fe$_{14}$Cu$_{5}$Mo$_{4}$/Fe$_{50}$Mn$_{50}$ (NiFeCuMo/FeMn) depend on NiFeCuMo thickness and substrate/buffer layer materials.  We used Ta and Cu as buffer layers, with both grown on the native oxide of Si (100) wafers, with Cu additionally grown onto a 140~nm thermal oxidize on Si (100).  Specifically, the Cu-buffered set had the structure Cu(300~$\textrm{\AA}$)/NiFeCuMo(90--300~$\textrm{\AA}$)/FeMn(150~$\textrm{\AA}$)/Ta(50~$\textrm{\AA}$), and were grown simultaneously on the two substrates.  The Ta-buffered set had the structure Ta(50~$\textrm{\AA}$)/NiFeCuMo(60--400~$\textrm{\AA}$)/FeMn(150~$\textrm{\AA}$), and were uncapped. The FeMn thickness of 150~$\textrm{\AA}$ was chosen so that the Blocking temperature of $\sim$400~K was independent of the antiferromagnet's thickness \cite{Bolon200754}. The substrates were ultrasonically cleaned in acetone and methanol for 5 minutes each,  blown dry with nitrogen gas, then inserting into the load lock. The samples were grown at ambient temperature in 3~mTorr of ultra high purity Ar in a magnetron sputtering system with a base pressure of 20~nTorr.  The compositions noted were those of the sputtering targets. All targets were presputtered for 10 minutes prior to deposition.  The sample holder was continually rotated during deposition, and the gun angle has been optimized to obtain deposition rates with variations less than 0.4\% over the entire 75~mm substrate holding plate.

\section{Results and Discussion}
X-ray diffraction confirms that both varieties of buffer induced (111) texturing in each sample variety.  The (111) orientation is specifically of interest because it is known to yield the largest exchange bias when using FeMn as the antiferromagnet \cite{Ritchie2002187}.  Each sample shows shifted (111) peaks relative to the bulk (43.2$^\circ$, 43.3$^\circ$, and 44.2$^\circ$ for FeMn, Cu, and NiFeCuMo, respectively, for Cu $k_{\alpha}$ radiation), indicating a reduction in lattice parameter along the growth direction. The (111) texture in the Ta-buffered NiFeCuMo is more coherent in the growth direction than that of the Cu-buffered samples, as indicated by the relative intensities of the peak near 44$^\circ$ (Fig.~\ref{xrd}).  Although the Ta seems to be the more promising buffer from this standpoint, a simple Scherrer analysis of the (111) peak indicates that the coherence length is only about half the film thickness.  Annealing or higher temperature deposition may improve the structure.  The origin of the weak (111) texturing in the Cu-buffered samples is not immediately obvious, since all the metals involved are FCC with only 2\% lattice parameter differences.  While still under investigation, we note that NiFeCuMo films may be susceptible to deposition-induced structural perturbations: we find it necessary to rotate the samples during growth in order to obtain reproducible magnetic properties; growing with the sputtering flux at a fixed angle relative to a stationary substrate leads to unexpected (and difficult to control) magnetocrystalline anisotropy.  It is possible that this structural sensitivity is playing a significant role in response to the differences in strain induced by the amorphous Ta and polycrystalline Cu buffers.

\begin{figure}
\begin{center}
\includegraphics[width=3.3in]{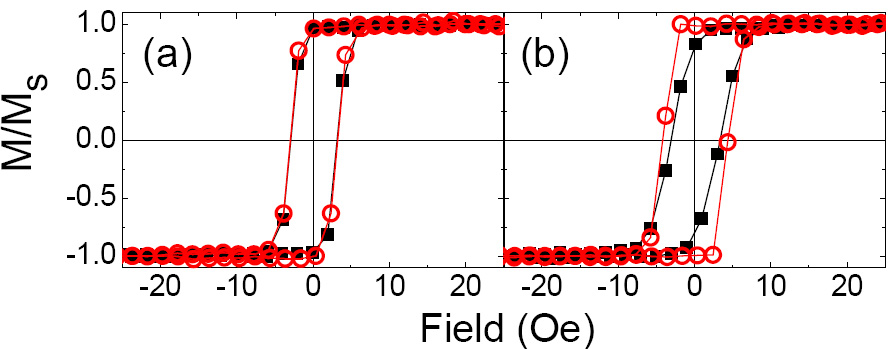}
     \caption{\small{(Color online) Hysteresis loops of  Cu(300~$\textrm{\AA}$)/NiFeCuMo(200~$\textrm{\AA}$)/Cu(300~$\textrm{\AA}$) samples grown simultaneously in (a) zero field, and (b) 250~Oe, as measured by VSM. The measurement field was applied parallel (red, open symbols) and perpendicular (black, solid symbols) to the deposition field direction.
     }} \label{controls}
\end{center}
\end{figure}
A custom substrate plate was used to deliver an in-plane field $\sim$250~Oe in a local region of the plate, while the opposite side of the plate has a field below the detection level of a calibrated Lakeshore 421 Gaussmeter.  This allows two control samples of Cu(300~$\textrm{\AA}$)/NiFeCuMo(200~$\textrm{\AA}$)/Cu(300~$\textrm{\AA}$) to be produced simultaneously, one with and one without an applied growth field \cite{5467356}.  X-ray reflectivity was used to confirm that the deposition rate was independent of the magnetic field used during deposition.   As shown in Fig~\ref{controls}(a), the sample deposited in zero field shows quite isotropic magnetic behavior, with no significant difference in hysteresis loop shape for the magnetization measured along two orthogonal directions; no measurable difference in M-H behavior was observed for any in-plane angle. In contrast, the field-grown sample has developed a uniaxial magnetic anisotropy, with the easy axis corresponding to the direction of the deposition field. The coercivity is slightly enhanced along the easy axis, while the hard axis coercivity is not measurably changed relative to the sample deposited in zero field.  The saturation fields are in line with previous results on NiFeCuMo thin films \cite{Egelhoff200690,jr.:013921}. These samples have no antiferromagnetic layer, and accordingly exhibit no exchange bias.

\begin{figure}
\begin{center}
\includegraphics[width=3.3in]{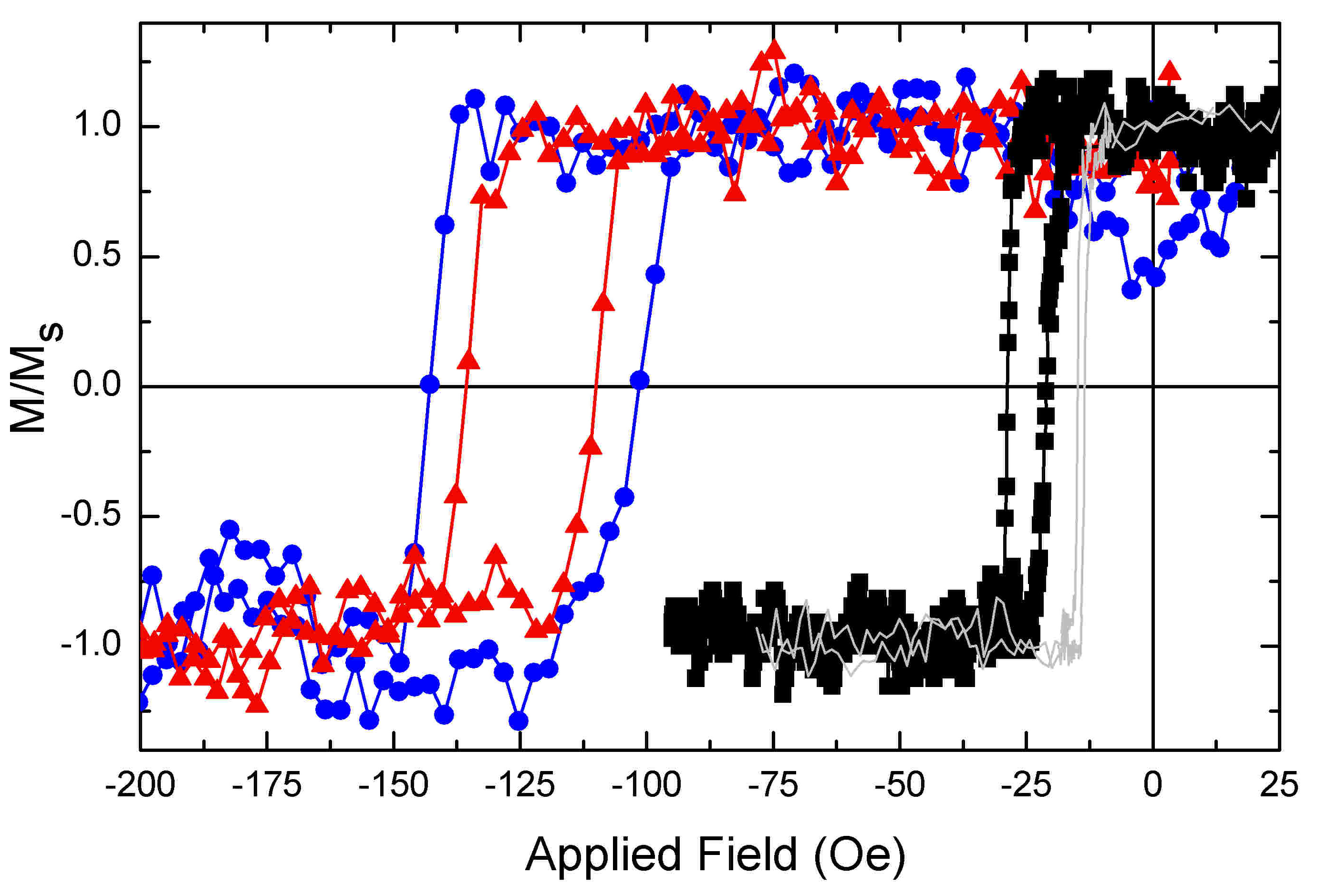}
     \caption{\small{(Color online) MH loops measured along the easy axes of NiFeCuMo(200~$\textrm{\AA}$)/FeMn(150~$\textrm{\AA}$) grown on substrate/buffer pairs of Si/Ta(50~$\textrm{\AA}$) (black squares), Si/Cu(300~$\textrm{\AA}$) (red triangles), SiOx/Cu(300~$\textrm{\AA}$) (blue circles). The MH loop of Ta/NiFeCuMo(400~$\textrm{\AA}$)/FeMn  (grey line) shows that soft magnetic properties can exist simultaneously with exchange bias in this system.
     }} \label{MH}
\end{center}
\end{figure}
Relative to the control samples, a clear exchange bias develops when FeMn is deposited in a field onto the NiFeCuMo.  Figure~\ref{MH} show room temperature hysteresis loops measured along the easy axes for 200~$\textrm{\AA}$ thick NiFeCuMo exchange biased with 150~$\textrm{\AA}$ FeMn.  Cu-buffered samples have significantly greater $H_{EB}$ than Ta-buffered samples. The coercive fields increase for all samples relative to the 4.3~Oe of the field-deposited control sample.  Si/Cu and SiOx/Cu samples have the most significant change, with  $H_C$~=~13~Oe and 21~Oe, respectively, while the Ta-buffered sample has $H_C$~=~9.1~Oe.

\begin{figure}
\begin{center}
\includegraphics[width=3.3in]{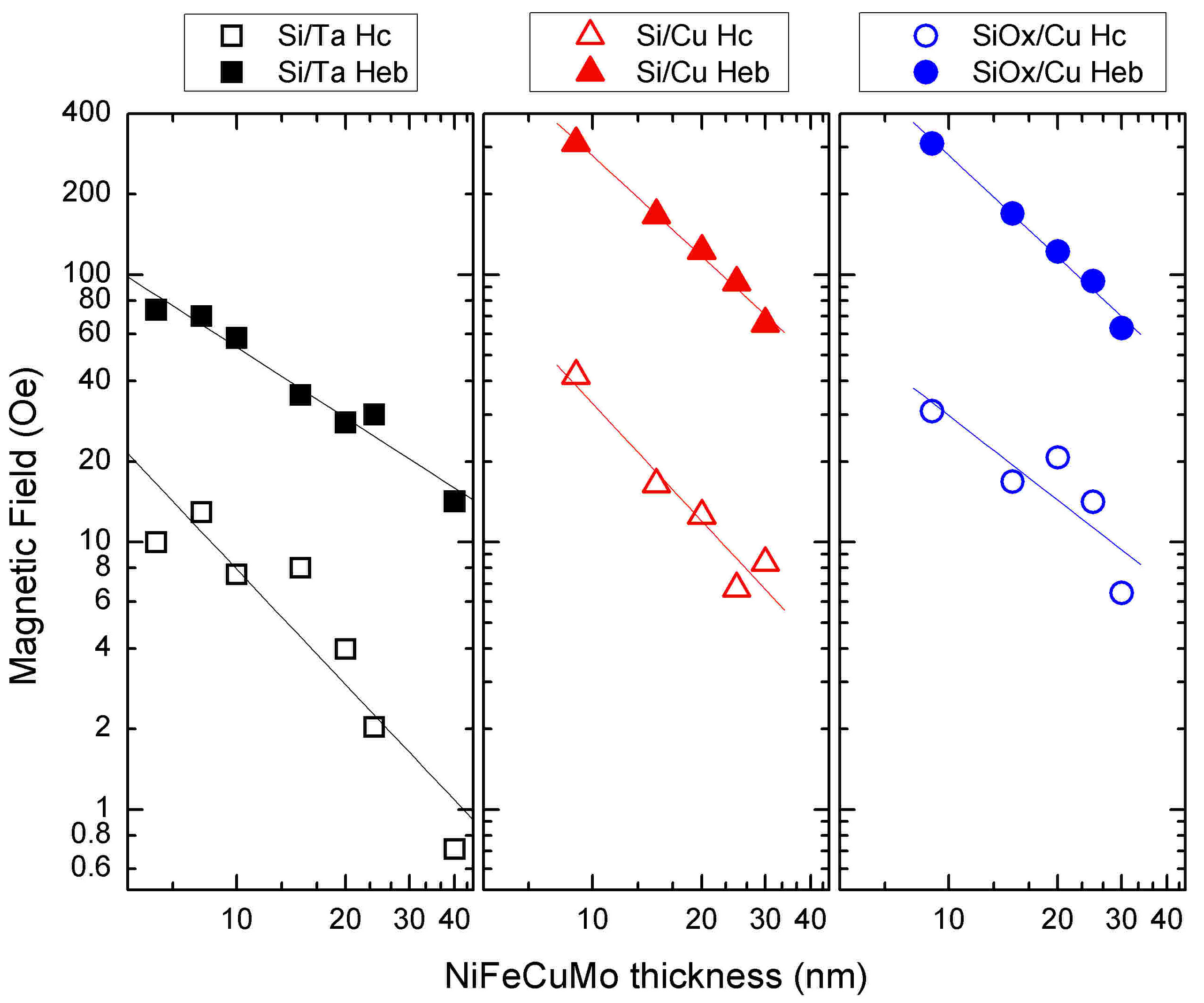}
     \caption{\small{(Color online) Exchange bias and coercive fields are inversely proportional to the ferromagnet thickness for all three sample sets; thin lines are linear fits.
     }} \label{eb-tfm}
\end{center}
\end{figure}
For a more global view, Fig.~\ref{eb-tfm} shows the thickness dependence of $H_{EB}$ and $H_C$ for each sample set.  Both $H_{EB}$ and $H_{c}$ are inversely proportional to the thickness, which is expected because exchange bias is an interface effect. The two Cu-buffered samples have nearly identical $H_{EB}$, suggesting that the interfaces in these samples are independent of the substrate.  Interestingly, the Ta-buffered samples have significantly lower  $H_{c}$ than the Cu-buffered samples for a given NiFeCuMo thickness.  With $H_{EB}$~=~14.1~Oe, $H_{c}~=~0.7$~Oe, and a saturation field on the order of 1~Oe, the Ta/NiFeCuMo(400~$\textrm{\AA}$)/FeMn sample shows that the soft magnetic properties of the mu-metal can be retained in exchanged biased structures.
\begin{figure}
\begin{center}
\includegraphics[width=3.3in]{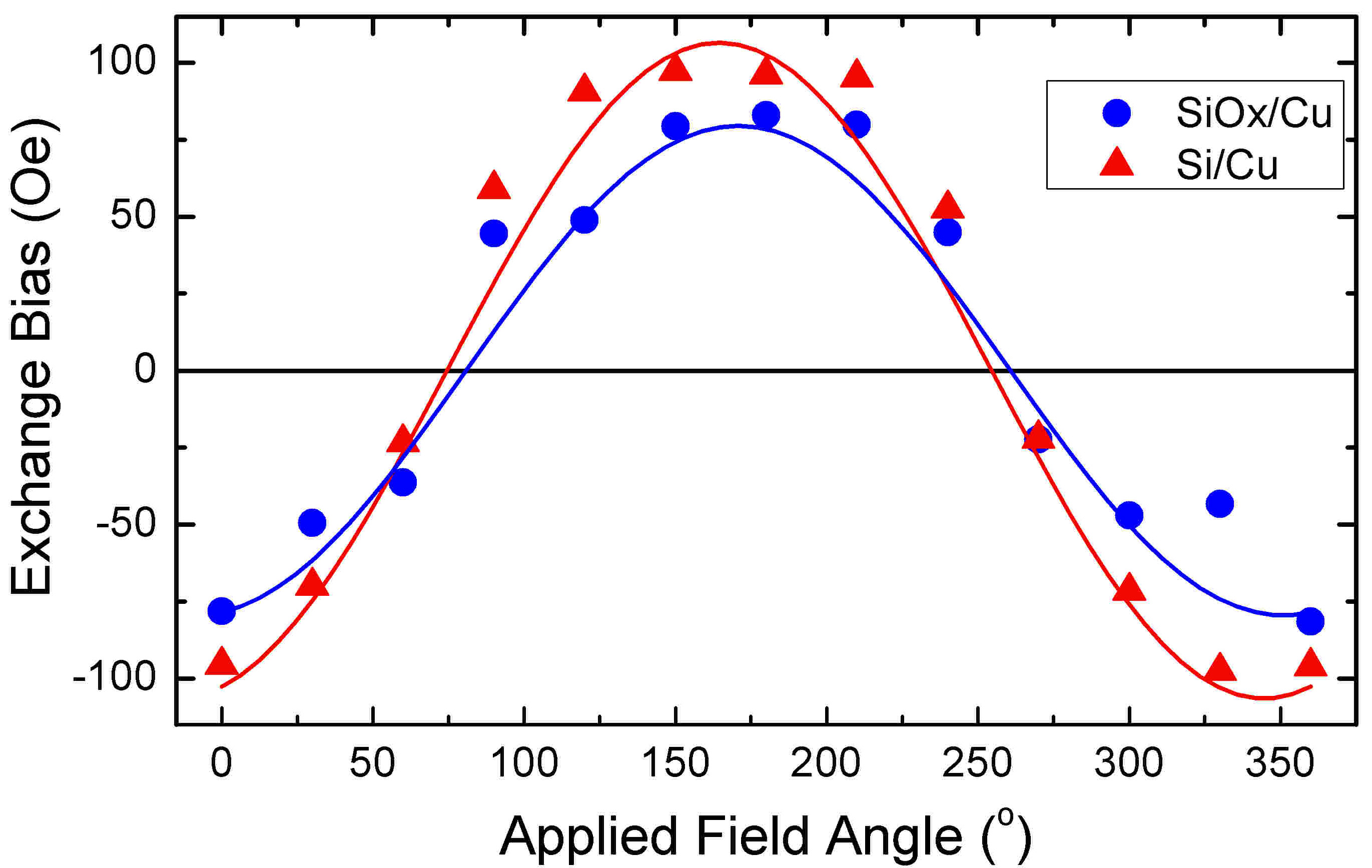}
     \caption{\small{(Color online) Exchange bias as a function of in-plane applied field angle for Cu(300~$\textrm{\AA}$)/NiFeCuMo(200~$\textrm{\AA})$/FeMn(100~$\textrm{\AA}$)/Ta(50~$\textrm{\AA}$) samples grown in 250~Oe onto Si (red triangles) and Si/SiOx (blue circles).  Fits to $H_{EB}\cos{\theta}$ yield $H_{EB}$ of -79~Oe and -106~Oe, respectively.
     }} \label{eb-theta}
\end{center}
\end{figure}
Figure~\ref{eb-theta} shows the exchange bias as a function of applied field angle relative to the deposition field direction for the 200~$\textrm{\AA}$ NiFeCuMo samples grown on Cu buffer layers\footnote{The uncapped Ta-buffered samples appear to have corroded over the course of two years with intermittent exposure to air, eliminating our ability to study their angular dependence; other measurements were performed within days of fabrication.}.  For both, the angular dependence of the exchange bias is fit well to  $H_{EB}\cos{\theta}$, with amplitudes of $-79$ and $-106$~Oe, respectively.

\begin{figure}
\begin{center}
\includegraphics[width=3.3in]{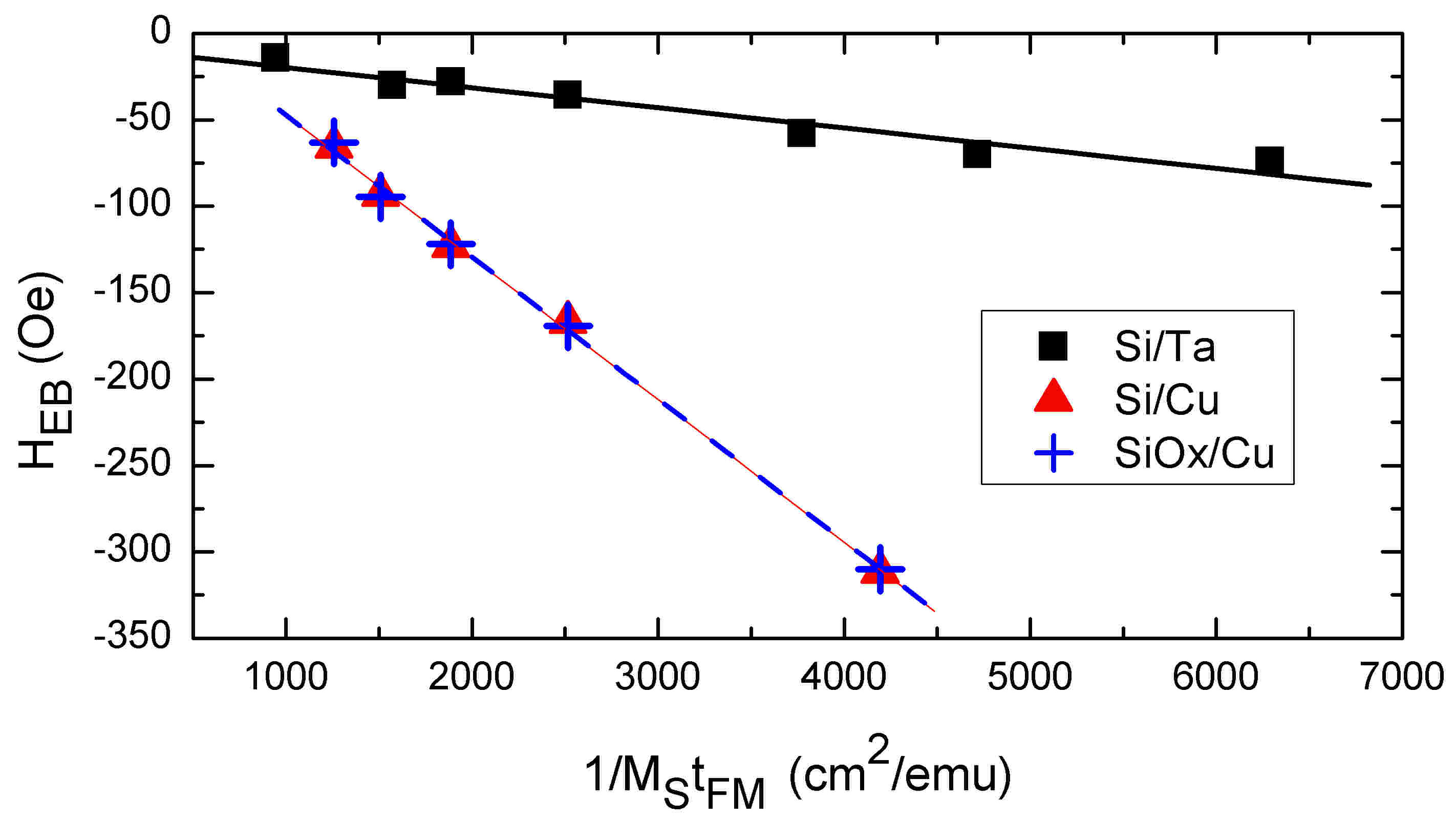}
     \caption{\small{(Color online) The interfacial exchange energy per unit area $J_{int}$ is the slope of the linear fits of the $H_{EB}$ vs 1/M$_s$t$_{FM}$, as described in the text.
     }} \label{jint}
\end{center}
\end{figure}

Using values of $H_{EB}$ the measured along the easy direction for each sample, we can determine the interfacial energy per unit area according to $J_{int} = M_st_{FM}H_{EB}$, where $M_s$ and $t_{FM}$ are the saturation magnetization (265~emu/cm$^3$) and thickness of the NiFeCuMo, respectively.  Figure~\ref{jint} shows that linear fits of the exchange bias as a function of $1/M_st_{FM}$ yield $J_{int}~=~-11.7\pm1.3$~merg/cm$^2$ for Si/Ta(5~nm), $-82.3\pm2.0$~merg/cm$^2$ for Si/Cu, and $-82.2\pm2.1$~merg/cm$^2$ for  SiOx/Cu.  These values are in agreement with previous energy densities using FeMn (111) as the antiferromagnet \cite{EBreviewIvan}.

\section{Conclusions}
Together, these results show that mu-metal exhibits classic exchange bias behavior when grown in contact with FeMn. The differences in magnetic properties between the Cu/Ni$_{77}$Fe$_{14}$Cu$_{5}$Mo$_{4}$/Fe$_{50}$Mn$_{50}$/Ta and Ta/Ni$_{77}$Fe$_{14}$Cu$_{5}$Mo$_{4}$/Fe$_{50}$Mn$_{50}$ samples are significant in respect to their applicability in low field sensing applications. The origin of the difference appears structural in nature.   Although both Cu and Ta lend themselves to (111) texturing of the NiFeCuMo and FeMn, samples with Ta buffers preserved the soft magnetic properties of the mu-metal most effectively. One notable result here is the ability to preserve the soft features of the mu-metal while inducing the unidirectional anisotropy.  This  may impact devices and structures employing soft magnetic materials, such as giant magnetoimpedance and related sensors, and exchange springs with tunable magnetization tilt angles.

\section{Acknowledgments}
This work was supported by the National Science Foundation. The Center for Integrated Functional Materials is supported by the USAMRMC. The authors thank Axel Hoffmann and John Pearson for assistance with the VSM.

%\bibliography{C:/BibFiles/career-ordered}

\end{document}